\documentclass[twocolumn,trackchanges]{aastex61}

\newcommand\aastex{AAS\TeX}

\accepted{24 July 2018}
\submitjournal{ApJ}

\shorttitle{\aastex\ Orbital properties of MSP-WD binaries}
\shortauthors{Hui et al.}

\begin{document}

\title{On the orbital properties of millisecond pulsar binaries} 

\correspondingauthor{C. Y. Hui}
\email{huichungyue@gmail.com,cyhui@cnu.ac.kr}

\author{C. Y. Hui}
\affil{Department of Astronomy and Space Science, Chungnam
National University, Daejeon 34134, Korea}

\author{Kinwah Wu}
\affiliation{Mullard Space Science Laboratory, University College London, Holmbury St.~Mary, Surrey, RH5 6NT, UK}

\author{Qin Han}
\affiliation{School of Astronomy and Space Science, Nanjing University, 163 Xianlin Avenue, Nanjing, Jiangsu 210093, China}
\affiliation{Mullard Space Science Laboratory, University College London, Holmbury St.~Mary, Surrey, RH5 6NT, UK}

\author{A. K. H. Kong}
\affil{Institute of Astronomy, National Tsing Hua University, Hsinchu, Taiwan}
\affil{Astrophysics, Department of Physics, University of Oxford, Keble Road, Oxford OX1 3RH, UK}

\author{P. H. T. Tam}
\affil{Institute of Astronomy and Space Science, Sun Yat-Sen University, Guangzhou 510275, China}


\begin{abstract}
We report a detailed analysis of the orbital properties of 
   binary millisecond pulsar (MSP) with a white dwarf (WD) companion. 
Positive correlations between the orbital period $P_{\rm b}$ and eccentricity $\epsilon$ are 
   found in two classes of MSP binaries with a He WD and with a CO/ONeMg WD, 
   though their trends are different. 
 The distribution of $P_{\rm b}$ is not uniform. Deficiency of sources 
  at $P_{\rm b}\sim35-50$~days (Gap 1) have been mentioned in previous studies. On the 
other hand, another gap at $P_{\rm b}\sim2.5-4.5$~days (Gap 2) is 
identified for the first time.  
Inspection of the relation between $P_{\rm b}$ and the companion masses $M_{\rm c}$ 
  revealed the subpopulations of MSP binaries with a He WD separated by Gap 1,  
  above which $P_{\rm b}$ is independent of $M_{\rm c}$ (horizontal branch)  
  but below which $P_{\rm b}$ correlates strongly with $M_{\rm c}$ (lower branch). 
Distinctive horizontal branch and lower branch separated by Gap 2 were identified  
  for the MSP binaries with a CO/ONeMg WD at shorter $P_{\rm b}$ and higher $M_{\rm c}$. 
Generally, $M_{\rm c}$ are higher in the horizontal branch than in the lower branch 
 for the MSP binaries with a He WD.  
These properties can be explained in terms of a binary orbital evolution scenario 
  in which the WD companion was ablated by a pulsar wind in the post mass-transfer phase. 
\end{abstract}



\section{Introduction} 
Shortly after the first millisecond pulsar (MSP) PSR~B1937+21 (Backer et al. 1982) was discovered,  
  a ``recycling" scenario for its formation, in which old neutron stars in binaries were spun up 
  by acquiring angular momentum through accreting material from a companion 
  (Alpar et al. 1982; Radhakrishnan \& Srinivasan 1982; Fabian et al. 1983), was proposed.   
The process would occur at the late evolutionary stage of low-mass X-ray binaries (LMXBs).  
The scenario explains the low surface magnetic fields, 
  the millisecond equilibrium rotational periods and the very low spinning-down rates 
  as observed in MSPs (Backer et al. 1983).   
It received further supports by the discovery of accreting millisecond X-ray pulsars 
  (e.g. SAX~J$1808.4-3658$, Wijnands \& van der Klis 1998) 
  and ``red-back" MSPs, which show alternating LMXB and rotation-powered states  
  (e.g. PSR~J$1023+038$, Archibald et al. 2009, 2010; Thorstensen \& Armstrong 2005; Takata et al. 2014). 
In spite of its success, 
   certain evolutionary aspects of MSPs, especially those in binaries, are yet to be satisfactorily explained 
   within the scenario's framework. 
Besides the initial distributions of the orbital separations and the companion masses, 
  how the progenitor systems had evolved through a common-envelope phase, 
  which has not been directly observed,  
  and how orbital angular-momentum was transported at various evolutionary stages 
  are still unclear (Tauris 1996; Taam et al. 2000). 
The complex evolution dynamics of MSP binaries is also reflected 
  a ``period gap" at $P_{\rm b}\sim23-56~{\rm days}$ 
  where there is a deficiency in source number (Tauris 1996; Taam et al. 2000).  
Its presence indicates a possible bifurcation process in operation,  
  causing divergent evolutionary paths for the subpopulations of the systems (Tauris 1996).

Theoretical investigations have predicted the relations between the orbital properties of MSPs. 
Two relations of the MSPs in a wide orbit of orbital period $P_{b}\gtrsim2$~days with companion mass 
$M_{\rm c}\lesssim2M_{\odot}$ have long been suggested as the dynamical fossils of the spin-up era 
(Phinney \& Kulkarni 1994). These are $P_{b} - M_{\rm c}$ relation
(Refsdal \& weigert 1971; Tauris \& Savonije 1999) and that between $P_{b}$ and the eccentricity $\epsilon$ (Phinney 1992). 
And hence, a thorough population analysis of the MSP binaries can help to 
to gain new insights into the intricate evolutionary paths of MSP binaries.   
With the expanded sample established by the recent observations, 
  we conducted a statistical analysis of orbital properties of MSP binaries with a WD companion.  
This article reports our findings and interpretations in the light of the new statistics. 


\section{Data Analysis}
We focus on the MSP binaries with a white-dwarf (WD) companion, 
   adopting the selection criteria: (i) the MSP rotational periods $P<40\;\!{\rm ms}$ 
  and (ii) the sources are in the Galactic field. 
MSP binaries in globular clusters (GC), which had involved different dynamical formation processes 
  (see Hui et al. 2010), were excluded. 
The system parameters of the sources were derived from the data in the updated ATNF pulsar catalog 
  (2018 April version; Manchester et al. 2005). 
Our first objective is to examine the relations among three key parameters: 
   the orbital period $P_{\rm b}$, 
   the orbital eccentricity $\epsilon$ and the mass of the WD companion $M_{\rm c}$.  
For the systems with precise measurements (cf. Table 2 in Ozel \& Freire 2016), we adopted their 
$M_{c}$ from the literature. 
For the others, we adopted an orbital inclination of $i=60^{\circ}$ 
   when deriving $M_{\rm c}$ from their mass functions by assuming a neutron star mass of $1.35M_{\odot}$.
To allow a constrained analysis, 
  we discarded the data with uncertainties $>50\%$ in $P_{b}$ and $\epsilon$.
The screening yielded a sample comprising of 58 MSP binaries with a helium (He) WD companion 
   and 25 MSP binaries with a carbon-oxygen/oxygen-neon-magnesium (CO/ONeMg) WD companion. 

We recognized that there are several recently discovered MSPs which fit our selection criteria 
are not included in the aforementioned sample. PSR~J2234+0611 (Antoniadis et al. 2016) and PSR~J1946+3417 (Barr et al. 2016) 
can be found in ATNF catalog but the nature of their companions are not specified in it. 
On the other hand, PSR~J1618-3921 (Octau et al. 2018) is not included in the ATNF catalog. 
The companions of all these systems appear to be He WDs. And their eccentricities lie in the range of 
$\epsilon\sim0.01-0.1$ which 
are higher than the general MSP population with He WD companions. Appending these systems to our sample, we have 61 MSP binaries 
with He WD companion for our analysis.

\begin{figure*}
\plotone{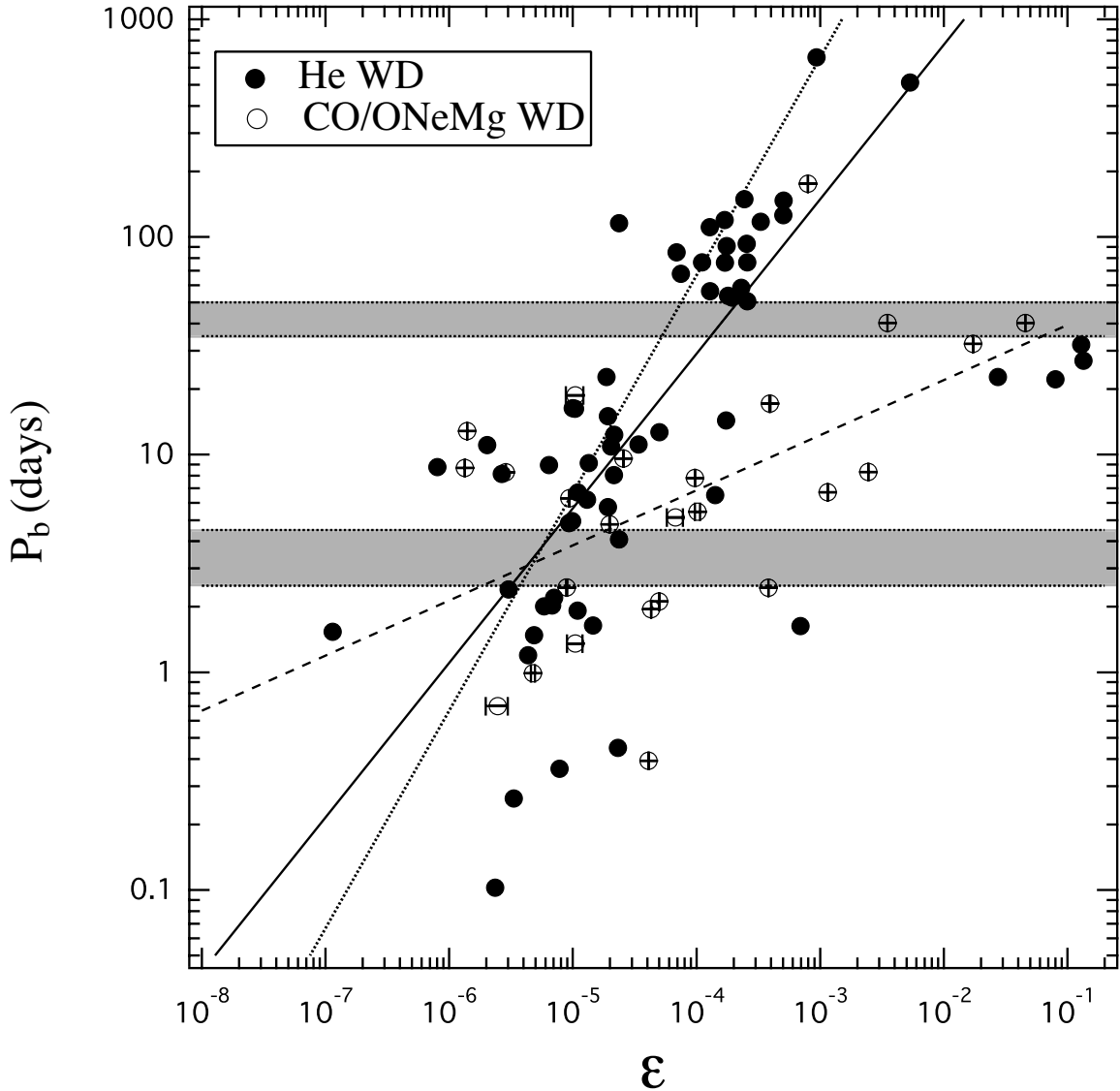}
\caption{The orbital period $P_{\rm b}$ and eccentricity $\epsilon$ relation 
  for MSP binaries with a He WD companion (solid symbols) 
   and with a CO/ONeMg WD companion (open symbols). 
The period gaps are marked as the shaded regions. 
The solid and dashed lines correspond to the best linear fits  
  between $\log P_{\rm b}$ and $\log\epsilon$ 
  for systems containing a He WD and a CO/ONeMg WD respectively. The simple relation predicted by Phinney (1992) has also been plotted for comparison 
(dotted line)}
\end{figure*}

\subsection{$\epsilon-P_{\rm b}$ relation}
Figure~1 shows the MSP binaries in the $\epsilon$-$P_{\rm b}$ plane.  
A parametric (linear correlation coefficient $r$) and a 
   non-parametric (Spearman rank correlation coefficient $\rho$) test were applied 
   for the correlation test between $P_{\rm b}$ and $\epsilon$, 
   giving $r=-3.85\times10^{-2}$ ($p$-value=0.77) and $\rho=0.72$ ($p$-value=$4.13\times10^{-11}$)  
   respectively for the MSP binaries with a He WD.  
The apparent discrepancy between the results 
  is caused by a group of outliers with $\epsilon >0.01$ (see Figure~1), which includes PSR~J1950+2414, 
PSR~J2234+0611, PSR~J1946+3417 and PSR~J1618-3921. 
The eccentricity abnormality is speculated to be due to an unusual event, 
   such as a delayed accretion-induced collapse of a massive WD 
   during the course of the system's evolution (see Freire \& Tauris 2014). 
Alternatively, Antoniadis (2014) proposes that such high eccentricity can be resulted from the dynamical interaction 
between the binary and a circumbinary disk over $\sim10^{4}-10^{5}$~yrs.

When these four eccentric binaries are excluded, the correlation analysis yields
  $r=0.69$ ($p$-value=$2.94\times10^{-9}$) and $\rho=0.74$ ($p$-value=$4.08\times10^{-11}$).   
Thus, both tests reconciled, concluding a strong $\epsilon$-$P_{\rm b}$ correlation. 
As $\rho$ is a non-parametric statistic, its estimate is therefore robust.  
The statistical significance it refers would be almost unaltered when the outliers are removed.  
A regression analysis excluding four eccentric MSPs yielded a relation  
\begin{equation}
  \log (P_{\rm b}/{\rm day})=(0.71\pm0.18)\log\epsilon + (4.30\pm0.81) 
\end{equation}
  for $P_{\rm b}$ and $\epsilon$. 
Here and hereafter unless otherwise stated    
the uncertainties of the parameters are of a $95\%$-confidence interval which are estimated by 
$t_{1-\alpha/2,\nu}SE$, where $t_{1-\alpha/2,\nu}$ is the $t$ statistic with $\alpha=0.05$ and a degree of freedom 
$\nu=$(no. of data point - number of free parameters) and $SE$ is the standard error of the corresponding parameters derived from the covariance matrix.

We also plotted the $\epsilon-P_{\rm b}$ relation predicted by Phinney (1992) 
($\epsilon\sim1.5\times10^{-4}\left(P_{\rm b}/100~{\rm days}\right)$; dotted line in Figure 1) for comparison. 
For $P_{b}\gtrsim1$~day, our best-fit based on the current sample (i.e. Equation 1) predicts a higher $\epsilon$ for 
a given $P_{b}$ than that suggested by Phinney (1992). 

For the MSP binaries with a CO/ONeMg (non-He) WD, there is no noticeable outlier. 
We obtained $r=0.19$ ($p$-value=0.37) and $\rho=0.45$ ($p$-value=$3.09\times10^{-2}$) 
  for the $\epsilon$-$P_{\rm b}$ correlation, 
  which has a weaker significance than that of the He WD case. 
The corresponding $\epsilon$-$P_{\rm b}$  relation is  
\begin{equation}
  \log (P_{\rm b}/{\rm day})=(0.25\pm0.18)\log\epsilon + (1.85\pm0.77)\ .
\end{equation}
The different trends in the $\epsilon$-$P_{\rm b}$  relations for the MSP binaries with a He WD and 
  with a non-He WD can be seen in the best-fit relations shown in Figure~1. 

\begin{figure*}
\plotone{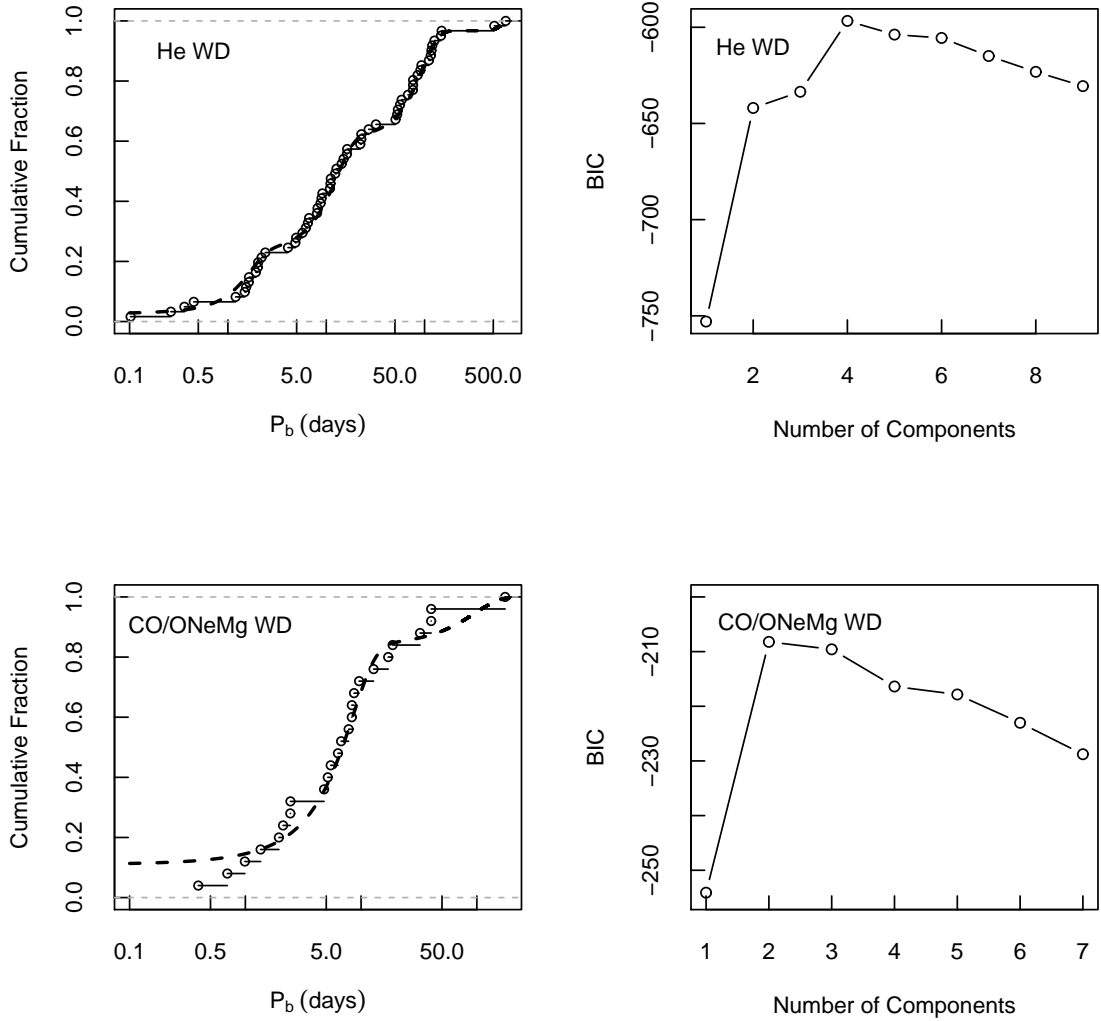}
\caption{
({\it Upper left panel}):
The unbinned empirical cumulative distribution function (CDF)
of the orbital periods of MSP binaries with a He WD companion in the Galactic field (open symbols). 
The dashed line illustrates four-component Gaussian mixtures model.
({\it Upper right panel}): The values of Bayesian information criterion (BIC) vs. the number of Gaussian components 
included in the model. 
({\it Lower left panel}): Same as upper left panel but for the MSPs with CO/ONeMg WD companions. 
({\it Lower right panel}): Same as upper right panel but for the MSPs with CO/ONeMg WD companions. 
The BIC values for eight and nine components models are undefined. 
}
\end{figure*}

\subsection{Period gaps}
Inspecting the orbital period distribution 
   revealed a deficiency of sources at  $P_{\rm b}\sim35-50~{\rm days}$  
   (the upper shaded region, Figure~1).   
Only two MSP binaries with a CO/ONeMg WD were there, but none with a He WD.   
We designate this period interval as ``Gap 1''.

If we limit ourselves to the systems follow $\epsilon-P_{\rm b}$ relation (i.e. ignore the outlying eccentric systems), 
the gap size can be relaxed to $P_{\rm b}\sim25-50~{\rm days}$.     
This gap is known (e.g. Taam et al. 2000; Tauris 1996; Camilo 1995). 
For those eccentric systems, their $P_{\rm b}$ appear to be rather similar (see Figure~1). Assuming the circumbinary disk scenario, 
Antoniadis (2014) has simulated the distribution of $P_{\rm b}$ from 1 day to 50 days which shows a jump of $\epsilon$ at $P_{\rm b}>10$~days. 
The author proposes that this might explain the existence of this period gap. At $P_{\rm b}\gtrsim50$~day, 
the $\epsilon-P_{\rm b}$ correlation can be recovered due to the cessation of hydrogen flashes for proto-WDs with mass $\gtrsim0.35M_{\odot}$ 
(Antoniadis 2014).

Besides ``Gap 1", we have identified another period gap, at $P_{\rm b}\sim2.5-4.5$~days.
We designate it as ``Gap 2", which is highlighted by the lower shaded region in Figure~1.
To examine the significance for the presence of these period gaps, we perform a model-based 
clustering on the distribution of $P_{\rm b}$ by using the \texttt{CRAN} {\it mclust} package 
(Fraley \& Raftery 2002, 2007). 

Since the binaries with He and CO/ONeMg WD companions have different evolutionary histories, 
we investigate their $P_{\rm b}$ separately. Assuming a mixture of Gaussian components, we
carried out the maximum likelihood fits. The calculation was repeated nine times with different 
number of components (1-9 Gaussians) included. The best model is chosen on the basis of Bayesian 
information criterion (BIC) values from the maximum likelihood estimations. 

The results are shown in Figure~2. For the $P_{\rm b}$ distribution of the binaries with a He WD, the BIC 
indicates that four Gaussian components best fit the data (see upper right panel of Figure~2). These four 
components are: 
\begin{equation}
\left\{
\begin{array}{lr}
\mathcal{N}_{1}(\mu_{1}=1.38~{\rm days}, \sigma_{1}=0.73~{\rm days}),& \left[0.20\right]\\
\mathcal{N}_{2}(\mu_{2}=10.35~{\rm days}, \sigma_{2}=5.75~{\rm days}),& \left[0.40\right]\\
\mathcal{N}_{3}(\mu_{3}=80.62~{\rm days}, \sigma_{3}=37.42~{\rm days}),& \left[0.37\right]\\
\mathcal{N}_{4}(\mu_{4}=590.56~{\rm days}, \sigma_{4}=78.52~{\rm days}), & \left[0.03\right]\\
\end{array}
\right.
\end{equation}

\noindent In Equation 3, $\mu$ and $\sigma$ are the mean and standard deviation for the corresponding Gaussian components. 
And the numbers in the square brackets are the fraction of the data covered by that component. 
The upper left panel of Figure~2 show the empirical cumulative distribution function (CDF) of the data for the 
$P_{\rm b}$ distribution of the binaries with a He WD (open symbols). Gap 1 and Gap 2 appears as the 
flattened regions in the CDF. The dashed curve is the four-component Gaussian mixture model. It appears that the model can 
reasonably describe the data. 

We further check if the separations among these components are significant by computing the Ashman's $D$ statistic 
(Ashman et al. 1994). All the pairs result in $D>12$ which indicate clear separations among them.  

Considering the clusters $\mathcal{N}_{1}$ and $\mathcal{N}_{2}$, the $1\sigma$ upper-bound of $\mathcal{N}_{1}$ 
and the $1\sigma$ lower-bound of $\mathcal{N}_{2}$ span a range of $\sim2.1-4.6$~days which encompasses Gap 2. 
Similarly, the clusters $\mathcal{N}_{2}$ and $\mathcal{N}_{3}$ are found to encompass Gap 1. 
Based on the aforementioned analysis,
the presence of these two period gaps are siginificant.

Although the BIC suggest the presence of $\mathcal{N}_{4}$ is also significant, this component only consists of two objects namely
PSR~J0407+1607 ($P_{\rm b}=669.07$~days) and PSR~J0214+5222 ($P_{\rm b}=512.04$~days). With such limited samples, we are not 
allowed to draw a firm conclusion for its existence and this component will not be considered in all subsequent analyses.

We apply the same analysis on the MSP binaries with CO/ONeMg WD companions. The values of the BIC (lower right panel of 
Figure~2) indicate that this population consists of two Gaussian components:

\begin{equation}
\left\{
\begin{array}{lr}
\mathcal{N}^{'}_{1}(\mu_{1}=5.95~{\rm days}, \sigma_{1}=4.67~{\rm days}),& \left[0.80\right]\\
\mathcal{N}^{'}_{2}(\mu_{2}=61.04~{\rm days}, \sigma_{2}=58.55~{\rm days}),& \left[0.20\right]\\
\end{array}
\right.
\end{equation} 

Comparing this model to the CDF constructed from the data (lower left panel of Figure~2), there 
are discrepancies between the model and the data at $P_{\rm b}\lesssim5$~days and $P_{\rm b}\gtrsim20$~days. 
The poor fit can be ascribed to the small sample of data. Although the analysis suggests this population 
might contain more than a single component, the location of both Gap 1 and Gap 2 cannot be constrained 
solely with the current MSP population with CO/ONeMg WD companions.


\begin{figure*}
\plotone{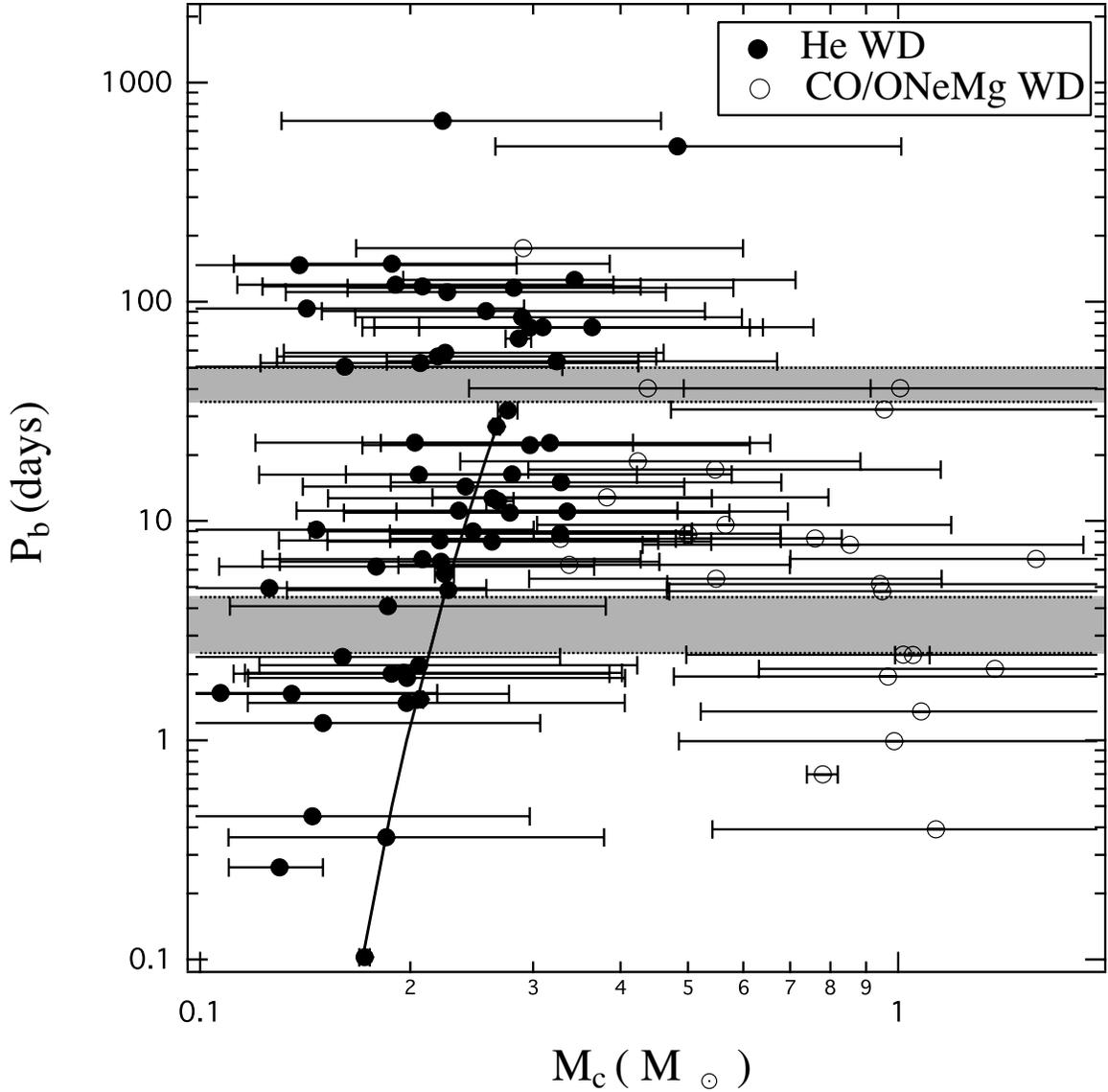}
\caption{The orbital period $P_{\rm b}$ and companion mass $M_{\rm c}$ relation 
  for MSP binaries with a He WD companion (solid symbols) 
  and with a CO/ONeMg companion (open symbols).
For those have been listed in the Table~2 of Ozel \& Freire (2016), which have 
dedicated measurement of $M_{\rm c}$, their values and the uncertainties are adopted. 
For the others, $M_{\rm c}$ are estimated from their mass functions by assuming an orbital inclination $i= 60^{\circ}$ 
and $M_{\rm NS}=1.35M_{\odot}$. Their error bars are estimated by varying $i$ from $18^{\circ}$ to $90^{\circ}$. 
The period gaps are marked as the shaded regions. The solid line illustrates the 
best-fit $M_{\rm c}$-$P_{\rm b}$ relation for the LB of systems with a He WD companion.}
\end{figure*}

\subsection{$M_{\rm c}$-$P_{\rm b}$ relation}
Figure~3 shows the $M_{\rm c}$-$P_{\rm b}$ distributions of the systems. 
Since their mass estimates adopted in this work are derived from the mass functions by 
assuming $M_{\rm NS}=1.35M_{\odot}$ and $i=60^{\circ}$, there is no error estimate provided by 
the ATNF catalog. For those sources with precise mass estimates, we adopted their values and 
the uncertainties given in Ozel \& Freire (2016). For the others, the uncertainties of $M_{\rm c}$ 
are difficult to be estimated. This can be ascribed to two facts: (1) the distribution of $i$ is 
likely to be uniform which leave $M_{\rm c}$ unconstrained without the measurement of $i$; 
(2) $M_{\rm NS}$ for MSPs can be different from the canonical value because of 
the accretion processes. Following van Kerkwijk et al. (2005), we reflect the uncertainties of $M_{c}$ by 
varying $i$ from $18^{\circ}$ to $90^{\circ}$ for those do not have dedicated mass measurement. The results are shown
in Figure~3. 

Interestingly, MSP binaries with a He WD are segregated into two subpopulations by Gap 1 
  whereas the MSP binaries with a non-He WD companion are divided into two subpopulations by Gap 2. 
For the MSP binaries containing a He WD above Gap 1 (the long-period systems),  
  a correlation analysis of $P_{\rm b}$ and $M_{\rm c}$ 
  gave $r=0.26$ ($p$-value $=0.25$) and $\rho=-0.05$ ($p$-value $=0.84$),  
  implying that $P_{\rm b}$ and $M_{\rm c}$ have no strong dependence. 
We refer this subpopulation as the ``horizontal branch" (hereafter HB) of the MSP binaries with a He WD companion. 
On the contrary, for the binaries below Gap 1 (the short-period systems)  
  there is a strong correlation between $P_{\rm b}$ and $M_{\rm c}$, 
  with $r=0.64$ ($p$-value $=6.74\times10^{-6}$) and $\rho=0.74$ ($p$-value $=4.37\times10^{-8}$). 
We refer this subpopulation as the ``lower branch" (hereafter LB) of the MSP binaries with a He WD companion.   
We note that all the eccentric systems follows the general trend of the LB. This is consistent with the 
prediction by Antoniadis (2014).

Tauris \& Savonije (1999) have computed the correlation between $M_{\rm c}$ and $P_{\rm b}$ numerically. They 
found that their model calculations can be fitted to a form of

\begin{equation}
\frac{M_{\rm c}}{M_{\odot}}=\left(\frac{P_{\rm b}}{b}\right)^{1/a}+c
\end{equation}

\noindent where $(a,b,c)$ depend on the composition of the donor and $P_{\rm b}$ in units of days.
Fitting Equation 5 to the LB of the MSP binaries with a He WD companion in our sample yields a set 
of parameters $a=4.91\pm2.26$, $b=(4.18\pm13.1)\times10^{5}$ and $c=0.12\pm0.04$. The results are comparable with Tauris \& Savonije (1999) (see Eq. 21 
in their paper). The best-fit $M_{\rm c}-P_{\rm b}$ relation is illustrated by the solid line in Figure~3.

From the data, we deduced that $(34\pm12)\%$ of MSP binaries 
  having a He WD companion would be in the HB and 
  $(66\pm12)\%$ in the LB. 
The 95\% confidence intervals were computed by the standard maximum likelihood Wald estimator.  

The binaries with a CO/ONeMg WD also show a HB (long-period systems) 
  and a LB (short-period systems) bisected by Gap 2  
  in the $M_{\rm c}$-$P_{\rm b}$  plane, 
  with also a turnover between the two branches.     
In comparison with the systems with a He WD, 
  they as a group are shifted to the lower right corner of the $M_{\rm c}$-$P_{\rm b}$ plane.  
This is partly because CO/ONeMg WDs are generally more massive.
It is, however, puzzling that systems with a CO/ONeMg WD 
  tend to have shorter orbital periods than the systems with a He WD.   
We deduced that there are $(68\pm18)\%$ and $(32\pm18)\%$ 
  MSP binaries containing a CO/ONeMg WD in their HB and the LB respectively.
No significant correlation was found 
  between $P_{\rm b}$ and $M_{\rm c}$ for both the populations above and below Gap 2.  

To examine the mass distributions of the WDs in the MSP binaries 
  we conducted 2-sample Kolmogorov-Smirnov (KS) and Anderson-Darling (AD) tests 
   for the following samples: 
  (a) WDs in all MSP binaries,  
  (b) He WDs in MSP binaries, and  
  (c) non-He WDs in MSP binaries, 
The SDSS field WDs in the Montreal White Dwarf Database (Dufour et al.~2017) [designated as (x)]. 
For the (x-a), (x-b) and (b-c) comparisons,  
   we obtained $p$-values $\ll 10^{-12}$ in both KS and AD tests; 
   for the (x-c) comparison, $p-$values of $1.1\times 10^{-4}$  in the KS test and $2.2 \times 10^{-7}$ in the AD test.   
A gaussian fit to the mass distribution of the SDSS WDs gave 
  $(M_{\rm wd}, \sigma) =$ (0.622, 0.157) [in units of M$_\odot$]
   while WDs in the MSP binaries gave $(M_{\rm c}, \sigma)=$ (0.394, 0.319).  
When MSP binaries were separated into the HB and LB sub-populations, 
  we obtained $(M_{\rm c}, \sigma)=$ (0.231, 0.070) [HB+LB], 
  (0.255,0.083) [HB] and (0.218,0.059) [LB] for the systems with a He WD  
  and $(M_{\rm c},\sigma) =$ (0.792, 0.342) [HB+LB], 
  (0.671,0.340) [HB] and (1.049,0.168) [LB] 
  for the systems with a CO/ONeMg WD.

\begin{figure*}
\plotone{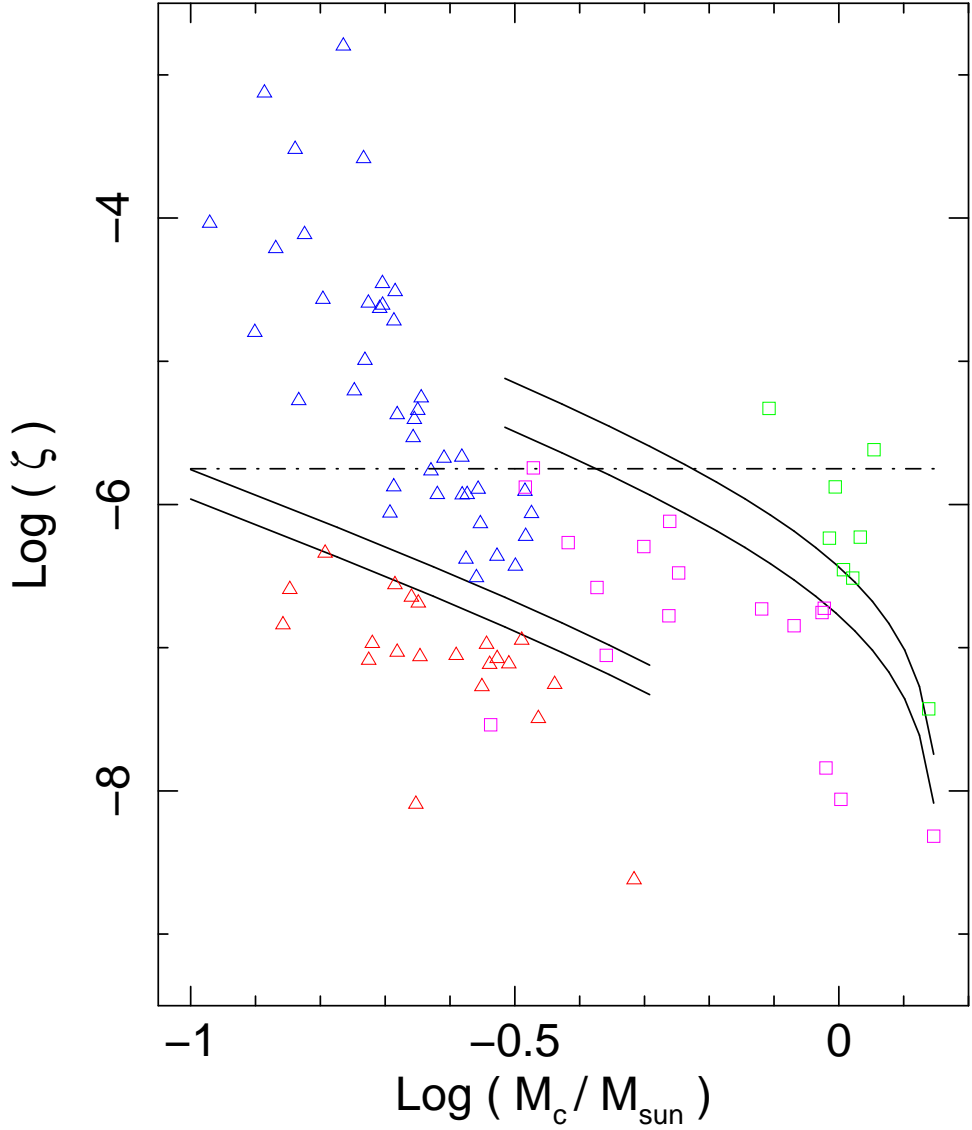}
\caption{The $M_{\rm c}$-$\xi$ relation for the MSP-WD binaries, 
  where $\xi\;\! (\equiv {R_{\rm c}}^2/4a^2M_{\rm c})$ is a measure of maximum amount of pulsar wind 
  that the WD can intercept. 
 The black curves are the period contours marking the 2 periods gaps, 
    with Gap 1 on the left side and Gap 2 on the right side in the $M_{\rm c}$-$\xi$ plane.  
 The orbital periods corresponding to the contours from top to bottom are therefore 
   $P_{\rm b}/{\rm day} = 2.5$, 4.5, 35 and 50. 
 MSP binaries with a He WD are represented by triangles 
   (red for HB systems and blue for LB systems), 
   and MSP binaries with a CO/ONeMg WD by squares
   (violet for HB systems and green for LB systems).  
The dashed line is a schematic approximate reference, 
  above which pulsar-wind ablation is able to cause significant mass loss from the WD. 
Orbital period evolves along the normals to the period contours,  
  and WD mass loss progresses along the horizontal axis. 
The ratio of the magnitude of the former to that of the latter is determined by $3[\alpha - f(q)]$ (see text).    
In our calculations, the MSPs are neutron stars  with 1.35-${\rm M}_\odot$, 
    and the WDs have radii given by the Nauenberg (1972) relation.
}
\end{figure*} 


\section{Discussions and Conclusions}   

In summary, our analyses have shown:   \\ 
\noindent
(i) The orbital periods $P_{\rm b}$ of the systems are not uniformly distributed, 
   with two gaps located at $P_{\rm b}\sim35-50$~days (Gap 1) and $2.5-4.5$~days (Gap 2).
    Gap 1 divides the MSP binaries with a He WD into two distinctive subgroups, the HB and LB, 
      and similarly Gap 2 separates the MSP binaries with a CO/ONeMg WD into the HB and LB subgroups.      \\ 
\noindent 
(ii) Both MSP binaries with a He WD and MSP binaries with a CO/ONeMg WD 
    show a positive correlation between $P_{\rm b}$ and $\epsilon$. 
  Their $\epsilon-P_{\rm b}$ relations are however not identical.  \\ 
\noindent 
(iii) Neither MSP binaries with a He WD nor those with a non-He WD 
  show $P_{\rm b}$-$M_{\rm c}$ dependence in the HB (above the period gap). \\ 
(iv) For the MSP binaries with a He WD, 
    $P_{\rm b}$ and $M_{\rm c}$ appear correlated in the LB (below Gap 1).  
   Such correlation is not present in the LB of the MSP binaries with a CO/ONeMg WD.  

These phenomena are consequence of the orbital evolutionary dynamics of the MSP-WD binaries, 
   which manifests as migration flows in the $M_{\rm c}$-$\xi$ plane (Figure 4).  
The variable $\xi \equiv {R_{\rm c}}^2/4a^2M_{\rm c}$, where $R_{\rm c}$ is the radius of the WD companion 
   and $a$ is the orbital separation of the system,    
  measures the amount of pulsar wind that the could be intercepted by the WD, per unit WD mass 
  and hence is an indicator of WD mass loss under pulsar-wind ablation.
In the $M_{\rm c}$-$\xi$ plane, 
  the rate of change in the system's orbital period are vectors normal to the constant $P_{\rm b}$ contours.  
The rate of change in the WD companion's mass are horizontal vectors with a negative direction. 
Adding these two vectors gives the individual migration velocity of the source, 
  whose horizontal component is always negative.  

The migration flow is driven by the angular momentum loss from the orbit and the mass loss from the companion star 
   caused by the interactions between the pulsar and the companion.   
The time derivative of Kepler's law gives the orbital evolutionary equation:  
  (with component stars of masses $m_1$ and $m_2$):  
\begin{eqnarray} 
 \frac{\dot J}{J}   & = &   
    \frac{\dot m_1}{m_1} \left[\frac{2+3q}{3(1+q)} \right] +  \frac{\dot m_2}{m_2} \left[\frac{3+2q}{3(1+q)}\right]   \cr
    & &   \hspace*{0.75cm}   + \frac{\dot P_{\rm b}}{P_{\rm b}} \left[ \frac{1}{3}\right] 
      - \frac{\dot \epsilon}{\epsilon} \left[ \frac{\epsilon^2}{(1-\epsilon^2)}\right]  \ , 
\end{eqnarray}  
  where $q = m_2/m_1$, $J$ is the orbital angular momentum, and ``$\cdot$'' denotes the time derivative. 
By setting ${\epsilon} \approx 0$  (almost circularized orbit), $m_2 = M_{\rm c}$, 
  ${\dot m}_1 = {\dot M}_{\rm MSP} \approx 0$ (insignificant mass gain) 
   and $f(q) = (3+2q)/3(1+q)$   
   and introducing a parametrization $(\dot J/J)/({\dot M}_{\rm c}/ M_{\rm c}) = \alpha(X_{\rm i},P,q) > 0$, 
   where $X_{\rm i}$ are variables intrinsic to the MSP,  
   we obtained the equation for period change in response to the companion's mass loss:    
\begin{eqnarray} 
  \frac{\dot P}{P} & \approx & \frac{\dot M_{\rm c}}{M_{\rm c}} \bigg[3 \big(\alpha -f(q) \big) \bigg] \ ,      
\end{eqnarray}    
Note that $f(q)$ is a slowly varying function, 
  with $f(q) \approx 0.977$ for $(M_{\rm MSP}, M_{\rm c}) = (1.35\;\! {\rm M}_\odot, 0.1\;\! {\rm M}_\odot)$ 
  and $f(q) \approx 0.830$ for $(M_{\rm MSP}, M_{\rm c}) = (1.35\;\! {\rm M}_\odot, 1.4\;\! {\rm M}_\odot)$.  
  
We propose that the period gaps were developed 
  when the progenitors of the MSP-WD binaries 
  were in the transition from being a mass-transfer system to a MSP system. 
By the time at this transition, the companion star's progenitor would have already shredded most of it mass, 
   implying the condition $q < 1$ being satisfied.   
This prevents a run-away mass transfer process and halts a rapid spiralling-in.   
If only mass exchange between the component stars occurs, $\dot J \approx 0$ on the dynamical timescale. 
Hence $\alpha \approx 0$ for the binary at this stage, when the mass transfer operates. 
The binary's orbit would expand in response to companion's mass loss. 
As such, the system migrates leftward and downward across the $M_{\rm c}$-$\xi$ plane. 
Pulsar-wind ablation will onset when the accretion pauses.  
The mass outflow from WD induced by the pulsar wind 
   and the viscous drag of the ablated material on the WD orbital motion 
   facilitates the angular-momentum extraction from the orbit.   
When the angular momentum loss is efficient, i.e. $\alpha \gg 1$, 
   mass loss from the companion will shorten the orbital period, 
   and the system migrate leftward and upward across the $M_{\rm c}$-$\xi$ plane.    
The period gaps are the separatrices that divides the upward and downward migration tracks of the sources 
  during the accretion-MSP transition. 

\noindent {\it (i) Locations of the two period gaps:}  
The progenitors of CO/ONeMg WDs 
  were able to evolve through the asymptotic giant branch and reach the horizontal giant branch, 
and they are  relatively massive (about 3~${\rm M}_\odot$ or higher).  
The MSP binaries with a CO/ONeNgWDs must have survived the complete spiralling-in  
  during the temperature oscillation phase of the companion star when C burning proceeds to O burning,  
  and the second common envelope phase (if present). 
The maximum size of the companion star is constrained by the orbital separation $a$, 
  and hence in the final mass-transfer episodes constrained by the period gap in our proposed scenario. 
With $a = 4.65\;\!{\rm R}_\odot (P_{\rm b}/{\rm day})^{2/3} (1+q)^{1/3}$,   
  $P_{\rm b} \sim (2.5-4.5)\;\! {\rm day}$ (Gap 2) corresponds to 
  $a \sim (8.6 - 13)(1+q)^{1/3}\;\!{\rm R}_\odot$, 
The radius of an evolved star with a 3-${\rm M}_\odot$ main-sequence progenitor 
   in the C/O burning stage will reach above $10\;\!{\rm R}_\odot$ 
  (see Maeder \& Meynet 1989),  
  consistent with the period-gap formation scenario (for Gap 2) that we propose,  
  if nuclear evolution drives the final episodes of the mass transfer process.  
He WDs have less massive main-sequence progenitors (about 1~${\rm M}_\odot$ or lower).
When the companion star of the progenitor MSP binary 
   evolve into the He burning red giant stage, it expands substantially.  
A common envelope could be avoided if the two stars have a sufficient large orbital separation.   
For these systems  mass transfer is expected to operate in a somewhat steady manner, 
  as the companion star is less massive than the neutron star.  
This leads to orbital expansion and period lengthening.  
$P_{\rm b} \sim (35-50)\;\! {\rm day}$ (Gap 1) corresponds to $a \sim (50 - 63)(1+q)^{1/3}\;\!{\rm R}_\odot$. 
The radius of a star, starting as 1-${\rm M}_\odot$ main-sequence star, is $\sim 30\;\!{\rm R}_\odot$ 
   at the end of its He burning (Charbonnel et al. 1999),    
  which $\sim 0.5$ times of the orbital separation inferred from Gap 1,   
  a configuration where Roche-lobe filling mass-transfer is possible 
  (see e.g. Eggleton 1983).   

\noindent {\it (ii) HB formation:} 
We interpret that the HB is a piling-up of systems, 
  caused by orbital expansion in the final mass-transfer episodes 
  when the progenitor binaries were at the transition from being an accretion system to a MSP system. 
The lacking of strong dependence of $M_{\rm c}$  
   for both MSP binaries with a He WD and with a non-He WD  
   is a consequence of the combination of followings: 
   (i) a weak dependence of $f(q)$ on $q$, 
   which gives ${\dot P_{\rm b}}/{\dot M_{\rm c}} \approx - 0.9 (P_{\rm b}/M_{\rm c})$, 
   and (ii) that the constant $P_{\rm b}$ contours are almost straight lines in the $M_{\rm c}$-$\xi$ plane  
   spanning from $\log (M_{\rm c}/{\rm M}_\odot) = -1.0$ to  $\log (M_{\rm c}/{\rm M}_\odot) \approx  0.0$.      
As such, the systems have a fairly uniform velocity over a wide $M_{\rm c}$ range   
  when migrating away from their respective period gaps in the $M_{\rm c}$-$\xi$ plane 
  on their course to become a ``full-fledged'' MSP binary. 
Although the systems would eventually evolve across the period gap later as MSP-WD binaries,  
  the process will be slow,  
  as at the HB the amount of pulsar wind intercepted by the WD is low (see Figure 4).  
Without a strong outflow from the WD, 
  direct extraction of angular momentum from the binary's orbit cannot be efficient.   
Moreover, there will be no viscous drag on the WD's motion when ambient material is absent.   
When $\alpha$ could not attain a high value, the MSP-WD binaries will linger in the HBs. 
    
\noindent {\it (iii) LB morphologies:}  
The pattern formation in the $M_{\rm c}$-$P_{\rm b}$ plot (Figure 2) 	
   is simply a reflection of the migration of the MSP binaries in the $M_{\rm c}$-$\xi$ plane (Figure 4), 
   which is driven by pulsar-wind ablation of the WD.  
The morphology of the LB of the MSP binaries with a He WD 
  is caused by the flow confluence of systems of all masses in the $M_{\rm c}$-$\xi$ plane, 
  in particular the rapid orbital evolution of the systems with a very low-mass WD ($\sim 0.1\;\!{\rm M}_\odot$). 
Low-mass WD have large radius. 
In addition to their efficient interception of pulsar wind, 
   they subject to large viscous drag if ambient material is present.  
For the lower WD-mass MSP binaries,  a large value for $\alpha$ can be attained,  
   implying a large $|{\dot P}_{\rm b}/P_{\rm b}|$ to $|{\dot M}_{\rm c}/M_{\rm c}|$ ratio.  
Thus, they have larger upward migration velocity component in the $M_{\rm c}$-$\xi$ plane 
  than their higher WD-mass counterparts.    
The confluence flow and the $M_{\rm c}$ dependent migration 
   explain the LB tilting for the MSP binaries with a He WD 
   and the apparent larger spread of $\xi$ in the LB at the low WD-mass end. 
Such a pattern is however not expected for the LB of the MSP binaries with a CO/ONeMg WD. 
The pulsar-wind ablation of their WD is not efficient enough 
  to drive rapid migration across $M_{\rm c}$-$\xi$ plane. 
At the high WD-mass end, the bending of the constant $P_{\rm b}$ contours 
  causes the ${\dot P}_{\rm b}/P_{\rm b}$ to have a strong horizontal projection 
  opposite to the direction of ${\dot M}_{\rm c}/M_{\rm c}$. 
Therefore, the massive-WD MSP systems can only migrate upward slowly in the $M_{\rm c}$-$\xi$ plane. 
Instead they slide gradually and only slightly deviate from the tangents of the constant $P_{\rm orb}$ contours.  
Only for systems with a WD of $\;\! \approx 1\;\!{\rm M}_\odot$ or lower, 
  such ``confinement'' to the migration flow becomes inefficient. 
Note that the MSP binaries with a low-mass CO/ONeMg WD in their HB 
  and the MSP binaries with a high-mass He WD in their LB  
  have very similar orbital periods, pulsar-wind ablation efficiencies and WD masses, 
  and hence subject to similar viscous drag.  
In the $M_{\rm c}$-$\xi$ plane,  
   MSP binaries with a low-mass CO/ONeMg WD in the HB 
   would therefore join the confluent flows of the MSP binaries with a He WD in LB 
   instead of migrating cross their own period gap, Gap 2 (see Figure 4).   
   
In summary, 
  we attributed the period gaps and their locations  
  to the conditions of the latest stages of stellar evolution of the WD progenitor. 
The evolutionary bifurcation of the MSP binaries with a He WD in the HB and LB 
  is due to relative efficiencies of angular momentum loss induced by the pulsar-wind ablation of the WD, 
  which naturally gives a positive correlation between $P_{\rm b}$ and $M_{\rm c}$ in the LB systems. 
The MSP binaries with a low-mass CO/ONeMg WD in the HB  
   have similar pulsar-wind ablation efficiencies as the MSP binaries with a high-mass He WD in the LB,  
   and hence these binaries migrate similarly in the $M_{\rm c}$-$\xi$ plane. 
The MSP binaries with a massive CO/ONeMg WD ($M_{\rm c}\gtrsim 1M_{\odot}$) 
  linger in the vicinity of their birth places  
  because the amount of pulsar wind intercepted by the WD is insufficient to drive a rapid orbital evolution. 
  
\appendix
\startlongtable
\begin{deluxetable}{llcr} 
\tablecaption{Orbital properties of MSP binaries with He WD companions.}
\tablehead{ 
\colhead{Pulsar Name\tablenotemark{a}} & \colhead{Orbital Period $P_{\rm b}$} & \colhead{Eccentricity $\epsilon$} & 
\colhead{Companion Mass $M_{\rm c}$} \\
                     & \colhead{(days)}    & \colhead{($10^{-6}$)}   & \colhead{($M_{\odot}$)}
}
\startdata
J0348+0432  &     0.1024($\pm7\times10^{-12}$) &  2.36($\pm1.0$)  &    0.172($\pm0.003$) \\
J0751+1807  &     0.2631($\pm7\times10^{-12}$) &  3.322($\pm0.5$) &     0.13($\pm0.02$) \\   
J1816+4510  &     0.3609($\pm2\times10^{-10}$) &  7.810($\pm2.0$)  &     0.185(-0.075,+0.194) \\ 
J1431-4715  &     0.4497($\pm7\times10^{-10}$) &  23.19($\pm0.8$)  &    0.145(-0.058,+0.152)  \\
J0613-0200  &     1.1985($\pm1.4\times10^{-11}$) &  4.350($\pm0.3$)  &     0.150(-0.060,+0.157) \\ 
J2043+1711  &     1.4823($\pm1.5\times10^{-11}$) &  4.868($\pm0.07$)  &     0.198(-0.081,+0.208)  \\
J1909-3744  &     1.5334($\pm1.3\times10^{-11}$) &  0.1140($\pm0.01$)  &     0.2067($\pm0.0019$)  \\
J0337+1715  &     1.6294($\pm5\times10^{-9}$) &   691.8($\pm0.2$)    &   0.135(-0.054,+0.142)  \\
J1622-6617  &     1.6406($\pm8\times10^{-9}$) &   14.56($\pm0.012$)    &   0.107(-0.042,+0.112)  \\
J1514-4946  &     1.9227($\pm5\times10^{-9}$) &   10.90($\pm0.003$)    &   0.198(-0.081,+0.208)  \\
J1902-5105  &     2.0118($\pm9\times10^{-10}$) &  5.864($\pm8\times10^{-7}$)    &   0.188(-0.076,+0.197)  \\
J0218+4232  &     2.0288($\pm9\times10^{-11}$) &  6.801($\pm0.4$)    &   0.196(-0.080,+0.206)  \\
J2017+0603  &     2.1985($\pm1.2\times10^{-10}$) & 7.060($\pm0.09$)   &    0.206(-0.084,+0.217)  \\
J1901+0300  &     2.3992($\pm6\times10^{-9}$) &  3.027($\pm1.0$)     &  0.160(-0.064,+0.168)  \\
J1045-4509  &     4.0835($\pm3\times10^{-10}$) & 23.67($\pm0.017$)     &  0.186(-0.075,+0.195)  \\
J0557+1550  &     4.8466($\pm4\times10^{-9}$) &  9.3($\pm0.4$)     &  0.227(-0.094,+0.240)  \\
J1745-0952  &     4.9435($\pm1.2\times10^{-8}$) & 9.849($\pm2.8$)    &   0.126(-0.050,+0.131)  \\
J0437-4715  &     5.7410($\pm4\times10^{-7}$) &   19.18($\pm0.0015$)    &   0.224($\pm0.007$)   \\
J1545-4550  &     6.2031($\pm8\times10^{-9}$) &  13.00($\pm0.4$)     &  0.179(-0.072,+0.188)  \\
J1017-7156  &     6.5119($\pm2\times10^{-6}$) &  142.0($\pm0.02$)     &  0.221(-0.091,+0.234)  \\
J1835-0114  &     6.6925($\pm4\times10^{-7}$) &  11.00($\pm3.0$)     &  0.208(-0.085,+0.220)  \\
J1543-5149  &     8.0608($\pm9\times10^{-9}$) &  21.46($\pm0.06$)    &  0.262(-0.110,+0.278)  \\
J1813-2621  &     8.1598($\pm1\times10^{-8}$) &  2.657($\pm1.0$)  &  0.220(-0.091,+0.233)  \\
J1125-6014  &     8.7526($\pm5\times10^{-8}$) &  0.8016($\pm0.15$) &  0.328(-0.141,+0.351)  \\
J1405-4656  &     8.9564($\pm7\times10^{-8}$) &  6.403($\pm2.5$)  &  0.246(-0.102,+0.260)  \\
J1056-7117  &     9.1388($\pm5\times10^{-7}$) &  13.42($\pm4.0$)  &  0.147(-0.059,+0.154)  \\
J1918-0642  &    10.9132($\pm1.6\times10^{-10}$) &  20.34($\pm1.5$)  &     0.278(-0.117,+0.295) \\ 
J1903-7051  &    11.0508($\pm2\times10^{-8}$) &  2.030($\pm0.005$)   &  0.336(-0.144,+0.359)  \\
J1804-2717  &    11.1287($\pm3\times10^{-9}$) &  34.06($\pm0.16$)    &  0.235(-0.097,+0.248)  \\
J1857+0943  &    12.3272($\pm1.8\times10^{-10}$) & 21.64($\pm0.03$)   &    0.267(-0.010,+0.014)   \\
J2236-5527  &    12.6892($\pm1.4\times10^{-7}$) &  50.20($\pm1.8)$   &    0.262(-0.110,+0.278)  \\
J1600-3053  &    14.3485($\pm3\times10^{-6}$) &  173.7($\pm0.009$)     &  0.240(-0.100,+0.254)  \\
J1810-2005  &    15.0120($\pm4\times10^{-8}$) &  19.24($\pm0.003$)     &  0.329(-0.141,+0.352)  \\
J1938+2012  &    16.2558($\pm1\times10^{-7}$) &  10.40($\pm0.9$)     &  0.206(-0.084,+0.217)  \\
J1741+1351  &    16.3353($\pm5\times10^{-10}$) & 9.984($\pm0.16$)     &  0.280(-0.118,+0.298)  \\
J1950+2414  &    22.1914($\pm1\times10^{-6}$) &  $7.981\times10^{4}$($\pm0.12$)     &  0.297(-0.126,+0.316) \\
J1709+2313  &    22.7119($\pm2\times10^{-8}$) &  18.70($\pm0.2)$     &  0.317(-0.136,+0.338)   \\
J1844+0115  &    50.6459($\pm1.1\times10^{-6}$) &  257.8($\pm1.2$)   &    0.161(-0.065,+0.169)   \\
J1825-0319  &    52.6305($\pm1.6\times10^{-6}$) &  193.9($\pm1.2$)   &    0.207(-0.085,+0.218)   \\
J0614-3329  &    53.5846($\pm8\times10^{-7}$) &  180.1($\pm0.1$)     &  0.324(-0.139,+0.347)   \\
J2033+1734  &    56.3078($\pm7\times10^{-8}$) &  128.7($\pm0.05$)     &  0.219(-0.090,+0.231)   \\
J1910+1256  &    58.4667($\pm8\times10^{-9}$) &  230.2($\pm0.018$)    &  0.225(-0.093,+0.237)   \\
J1713+0747  &    67.8251($\pm1.6\times10^{-9}$) & 74.94($\pm0.0006$)    &   0.286($\pm0.012$)    \\
J1455-3330  &    76.1746($\pm1.1\times10^{-8}$) & 169.6($\pm0.013$)    &   0.297(-0.126,+0.316)   \\
J1125-5825  &    76.4032($\pm5\times10^{-8}$) &  257.2($\pm0.03$)     &  0.310(-0.132,+0.330)   \\
J2019+2425  &    76.5116($\pm2\times10^{-8}$) &  111.1($\pm0.04$)     &  0.364(-0.158,+0.392)   \\
J1850+0124  &    84.9499($\pm4\times10^{-6}$) &  69.00($\pm1.3$)     &  0.289(-0.122,+0.308)   \\
J1935+1726  &    90.7639($\pm2\times10^{-5}$) &  175.8($\pm4.0$)     &  0.257(-0.107,+0.272)   \\
J2229+2643  &    93.0159($\pm1.5\times10^{-7}$) & 255.3($\pm0.04$)   &   0.142(-0.057,+0.149)   \\
J1751-2857  &   110.7465($\pm4\times10^{-8}$) &  127.9($\pm0.03$)     &  0.226(-0.093,+0.239)   \\
J1853+1303  &   115.6538($\pm1.4\times10^{-8}$) & 23.69($\pm0.006$)    &   0.281(-0.119,+0.299)   \\
J1955+2908  &   117.3491($\pm6\times10^{-8}$) &  330.2($\pm0.018$)   &  0.208(-0.085,+0.220)   \\
J1529-3828  &   119.6748($\pm1.6\times10^{-5}$) & 168.6($\pm1.4$)    &   0.191(-0.078,+0.201)   \\
J2302+4442  &   125.9353($\pm1.3\times10^{-7}$) & 503.0($\pm0.017$)    &   0.344(-0.148,+0.369)   \\
J1643-1224  &   147.0173($\pm7\times10^{-5}$) &  505.8($\pm0.009$)     &  0.139(-0.055,+0.145)   \\
J1708-3506  &   149.1332($\pm4\times10^{-7}$) &  244.5($\pm0.1$)      &  0.188(-0.076,+0.198)   \\
J0214+5222  &   512.0397($\pm3\times10^{-4}$) &  $5.328\times10^{3}$($\pm0.5$)  &  0.483(-0.218,+0.528)   \\
J0407+1607  &   669.0704($\pm1\times10^{-4}$) &  936.8($\pm0.6$)     &  0.223(-0.092,+0.235)   \\
J2234+0611  &    32.0014($\pm1\times10^{-7}$) &  $1.293\times10^{5}$($\pm0.014$)     &  0.276($\pm0.009$)    \\
J1946+3417  &    27.0199($\pm5\times10^{-8}$) &  $1.345\times10^{5}$($\pm0.017$)     &  0.2659($\pm0.003$)   \\
J1618-3921  &    22.7455($\pm1.9\times10^{-7}$) &  $2.741\times10^{4}$($\pm1.0$)   &    0.2030(-0.083,+0.214)   \\
\enddata
\tablenotetext{a}{The errors of $M_{\rm c}$ for the pulsars marked with * are adopted from van Kerkwijk et al. (2005) and references therein. For the others, assuming the pulsar mass to be $1.35M_{\odot}$,  
their errors are estimated from the mass functions with the inclination angles vary from $i=90^{\circ}$ to 
$i=18^{\circ}$. }
\end{deluxetable}

\newpage
\begin{deluxetable*}{llcr}
\tablecaption{Orbital properties of MSP binaries with CO/ONeMg WD companions.}
\tablehead{
\colhead{Pulsar Name\tablenotemark{a}} & \colhead{Orbital Period $P_{\rm b}$} & \colhead{Eccentricity $\epsilon$} &
\colhead{Companion Mass $M_{\rm c}$} \\
                     & \colhead{(days)}    & \colhead{($10^{-6}$)}   & \colhead{($M_{\odot}$)}
}
\startdata
J1952+2630  &     0.3919($\pm7\times10^{-11}$)  &   40.85($\pm0.1$)   &    1.133(-0.591,+1.379)\\ 
J1802-2124  &     0.6989($\pm5\times10^{-12}$)  &    2.474($\pm0.5$)  &     0.78($\pm0.04$)  \\
J1525-5545  &     0.9903($\pm7\times10^{-10}$)  &    4.754($\pm0.17$)  &     0.987(-0.502,+1.172) \\
J1435-6100  &     1.3549($\pm1.8\times10^{-9}$) &    10.47($\pm1.5$)  &     1.079(-0.558,+1.301) \\
J1949+3106  &     1.9495($\pm2\times10^{-6}$)   &    43.06($\pm0.07$)  &     0.967(-0.490,+1.143) \\
J1439-5501  &     2.1179($\pm3\times10^{-9}$)  &    49.85($\pm1.5$)  &     1.376(-0.744,+1.749) \\
J2222-0137  &     2.4458($\pm7\times10^{-11}$) &    381.0($\pm0.03$)  &     1.05($\pm0.06$)  \\
J2053+4650  &     2.4525($\pm2\times10^{-10}$) &    8.900($\pm0.1$)  &     1.017(-0.520,+1.212)\\
J1337-6423  &     4.7853($\pm5\times10^{-9}$)  &    19.85($\pm0.09$)  &     0.949(-0.479,+1.119)\\
J1933+1726  &     5.1539($\pm2\times10^{-8}$)  &    67.45($\pm10.0$)  &     0.942(-0.475,+1.109)\\
J0721-2038  &     5.4608($\pm8\times10^{-8}$)  &    102.0($\pm5.0$) &     0.548(-0.252,+0.605)\\
J1603-7202  &     6.3086($\pm5\times10^{-10}$) &    9.338($\pm0.005$)&     0.338(-0.146,+0.362)\\
J1227-6208  &     6.7210($\pm4\times10^{-9}$)  &    $1.149\times10^{3}$($\pm3.0$)  &     1.576(-0.875,+2.076)\\
J1022+1001  &     7.8051($\pm1.1\times10^{-6}$) &    97.23($\pm0.014$)  &     0.853(-0.422,+0.989)\\
J1943+2210  &     8.3115($\pm2\times10^{-8}$)  &    2.865($\pm0.8$)  &     0.3282(-0.141,+0.351)\\
J0621+1002  &     8.3187($\pm3\times10^{-7}$)   &    $2.457\times10^{3}$($\pm0.07$)  &     0.76(-0.28,+0.07) \\ 
J1614-2230  &     8.6866($\pm7\times10^{-11}$)  &    1.336($\pm0.007$)  &     0.5($\pm0.006$)  \\ 
J1101-6424  &     9.6117($\pm3\times10^{-7}$)   &    25.68($\pm1.3$)  &     0.566(-0.262,+0.626)\\
J1933-6211  &    12.8194($\pm8\times10^{-10}$)  &    1.397($\pm0.04$)  &     0.382(-0.167,+0.412)\\
J1750-2536  &    17.1416($\pm4\times10^{-6}$)  &    392.0($\pm4.0$)  &     0.547(-0.252,+0.603)\\
J0900-3144  &    18.7376($\pm9\times10^{-10}$) &    10.49($\pm1.7$)  &     0.424(-0.188,+0.459)\\
J2045+3633  &    32.2978($\pm1\times10^{-6}$)  &    $1.721\times10^{4}$($\pm0.05$) &     0.955(-0.483,+1.127)\\
J1420-5625  &    40.2945($\pm4\times10^{-6}$)  &    $3.500\times10^{3}$($\pm3.0$)  &     0.438(-0.195,+0.475)\\
J1727-2946  &    40.3077($\pm3\times10^{-8}$)  &    $4.563\times10^{4}$($\pm0.16$) &     1.007(-0.514,+1.199)\\
J1640+2224  &   175.4607($\pm7\times10^{-9}$)  &    797.3($\pm0.013$)  &     0.290(-0.123,+0.309) \\
\enddata
\tablenotetext{a}{The errors of $M_{\rm c}$ for the pulsars marked with * are adopted from van Kerkwijk et al. (2005) and references therein. For the others, assuming the pulsar mass to be $1.35M_{\odot}$, 
their errors are estimated from the mass functions with the inclination angles vary from $i=90^{\circ}$ to
$i=18^{\circ}$.}
\end{deluxetable*}
    

\acknowledgments
CYH is supported by the National Research Foundation of Korea grant 2016R1A5A1013277; 
AKHK by the Ministry of Science and Technology of Taiwan
    grants 106-2918-I-007-005 and 105-2112-M-007-033-MY2; 
PHTT by the SYSU One Hundred Talents Program;  
   and QH by a NJU School of Astronomy and Space Sciences Overseas Research Scholarship 
    and a UCL-MSSL Summer Research Studentship.  


\end{document}